\documentclass[twocolumn,preprintnumbers]{revtex4}
\usepackage{amssymb}
\usepackage{amsmath}
\usepackage{multirow}
\usepackage{lineno}
\usepackage{graphicx}
\usepackage{bm}
\usepackage{soul}
\usepackage{fancyhdr}
\usepackage[colorlinks=true,citecolor=blue,linkcolor=blue]{hyperref}
\usepackage[usenames]{color}
\usepackage{verbatim}
\usepackage{float}

\begin{document}

\title{Highly-efficient terahertz radiation generated by surface electrons from laser-foil interactions}

\author{Ke Hu}
\affiliation{Department of Physics, Chalmers University of Technology,
	41296 Gothenburg, Sweden}
\author{Longqing Yi}
\thanks{longqing@chalmers.se}
\affiliation{Department of Physics, Chalmers University of Technology,
	41296 Gothenburg, Sweden}

\date{\today}

\begin{abstract}
A novel scheme for generating powerful terahertz (THz) radiation based on laser-solid interactions is proposed.
When a $p$-polarized femtosecond laser  impinges obliquely on a plane solid target and the target partially blocks the laser energy, surface electrons are extracted out and accelerated by the laser fields, forming a low-divergence electron beam.
A half-cycle THz radiation pulse is emitted simultaneously as the beam passes by the edge of the target, due to coherent diffraction radiation.
Our particle-in-cell simulations show that
the relativistic THz pulse can have an energy of a few tens of millijoule and the conversion efficiency can be over 1$\%$ with existing $\sim$J level femtosecond laser sources.
\end{abstract}


\maketitle

Powerful terahertz radiation sources have attracted considerable attention due to its usage in many fields of science \cite{Hoffmann2011, Dhillon2017, Tonouchi2007}, such as THz spectroscopy of condensed matter or biological issues \cite{Siegel2004, Pickwell2006, Globus2003}, nonlinear THz optics \cite{Hebling2008} and resonant control of materials \cite{Kampfrath2013, Matsunaga2012}.
Conventional laser-based THz radiation sources
include optical rectification from
nonlinear crystals \cite{Hirori2011, Shalaby2015} and switched photoconducting antennnas \cite{Weling1994}.
The peak fields are limited at the order of $1$ or $2$ MV/cm and the radiation energies are smaller than $100\ {\rm \mu J}$.
Another option, accelerator driven sources, can produce THz radiation with higher electromagnetic fields ($>10$ MV/cm) and higher energies ($>600\ {\rm \mu J}$) \cite{Wu2013}.
However, they require linacs or storage rings to accelerate ultrashort relativisic electron bunches.
The low accessibility of such large-scale, expensive facilities hinder the broad research on this approach.

Recently, laser-plasma interactions have been considered as
a new method to produce strong THz radiation \cite{Leemans2003,Xie2006}.
When a solid foil is irradiated by a pump laser with intensities over $10^{18}\ {\rm W/cm^2}$, ultrafast electron bunches are produced, which lead to THz radiation emitted in both forward and backward directions \cite{Hamster1993, Gopal2013}. The forward emission is mostly attributed to coherent transition radiation (CTR) emitted by
a portion of hot electrons moving forward and crossing the rear surface \cite{Schroeder2004,Liao20192}. As for the backward THz radiation, two major mechanisms have been proposed.
The first one also relies on CTR of the backward moving electrons that is transmitted through the front plasma-vacuum boundary.
The second mechanism, also known as the antenna model, is attributed to the lateral current within the region of low-density plasma on the front surface, confined by the electrostatic fields \cite{Sagisaka2008}.

Several experimental and numerical studies have shown that the energy of fast electrons in laser-plasma interactions is normally on the order of $\sim100\ {\rm keV}$ or $\sim{\rm MeV}$ \cite{Li2016,Ding2013} , so, most of the hot electrons in the low-energy end of spectrum do not contribute to THz generation as they cannot escape the electrostatic fields near the target bulk.
Also the large beam divergence observed in the experiments suppresses conversion efficiency \cite{Schroeder2004},
and the peak THz amplitude is below $1$ GV/cm \cite{Ding2016, Liao2016, Liao20191}.
The strongest THz radiation in laser-foil interactions reported in experiments by far is above the millijoule level \cite{Liao20192}. However, the total conversion efficiency is smaller than $10^{-3}$.

In this letter, we report that, when a pump laser  impinges obliquely on a solid foil, well-collimated surface electrons are produced, which can serve as sources to generate strong THz radiation when passing by the edge of the target.
Those electrons are extracted out from the front side of the target and accelerated mainly by the electromagnetic fields of the incident laser pulse.
They have favourable features such as high charge (several nC), relatively small divergence ($\sim20^{\circ}$) and large energy (a few MeV).
Based on the mechanism of coherent transition radiation, such features can lead to a peak THz amplitude of a few ${\rm GV/cm}$, an energy of tens of mJ, and a conversion efficiency around $1\%$.

\begin{figure*}[htb]
	\centering
	\includegraphics[width=1.0\textwidth]{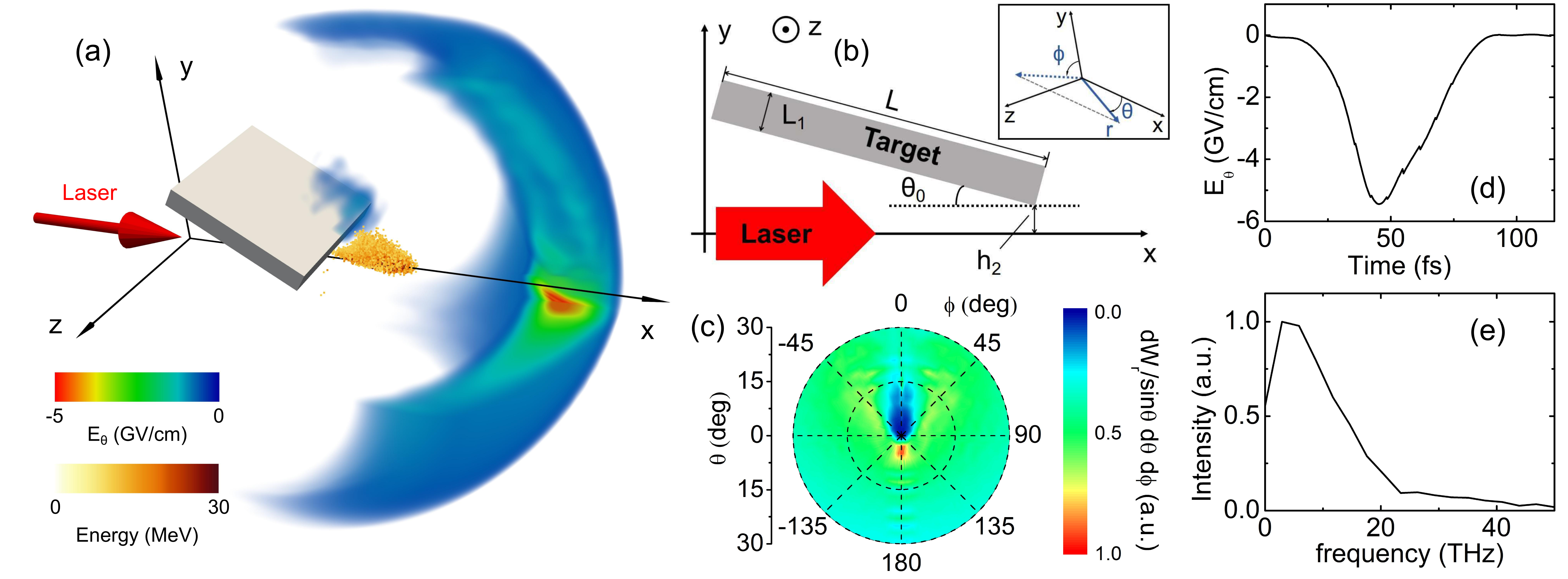}
	\caption{
		(a) 3D schematic setup of the proposed scheme.
		The polar component of the THz electric field $E_{\theta}$ (frequency $f < 60\ {\rm THz}$) at simulation time $t=330\ {\rm fs}$ is shown with rainbow colourscale.
		Nearly a quarter of the fields ($y>-3\ {\rm \mu m}$ and $z>0\ {\rm \mu m}$) is removed to display the intensity inside.
		The orange dots are fast electrons ($\gamma>20$) at $t=200\ {\rm fs}$ and the color represents their energy.
		(b) 2D schematic setup of the proposed scheme view in the $x-y$ plane and  demonstration of the spherical coordinate system (inset).
		(c) The angular distribution of radiated THz energy at $t=333\ {\rm fs}$.
		(d) 1D THz radiation field $E_{\theta}$ observed at $\theta=-5.6^{\circ}$ and $\phi=180^{\circ}$.
		(e) Spectrum of the radiation field shown in (d).
	} 		
\end{figure*}

The three-dimensional (3D) schematic setup is illustrated in Fig.~1(a) and the definitions of important parameters are shown in Fig.~1(b). A $p$-polarized incident laser pulse propagates along the $x$ axis, partially blocked by a solid foil.
The front surface of the foil is tilted $\theta_0=30^\circ$ with respect to the laser axis, and it is placed such that its right edge is $x_0=30\ {\rm \mu m}$ on the $x$ axis, and $h_2=2\ {\rm \mu m}$ above the $x$ axis.

When the laser arrives, the surface electrons are extracted and form compact beams in the laser field, which give rise to strong diffraction radiation as they pass by the right edge of the foil.
The laser pulse has an intensity of $I=1.37\times 10^{20}$ W/cm$^2$, which corresponds to a normalised vector potential $a_0=eE_0/m_ec\omega_0=10$, where $E_0$ is the amplitude of laser electric field, $m_e$ is the mass of an electron, $c$ is the speed of light and $\omega_0=2\pi c/\lambda_0$ is the laser frequency, with $\lambda_0=1\ {\rm \mu m}$ the wavelength. The laser beam has a Gaussian-shaped temporal profile, with FWHM duration of $T=35\ {\rm fs}$ and a spot size of $w_0=4\ {\rm \mu m}$.
We have assumed the length of the foil ($L$) is great enough so that its left edge does not touch the laser field. The thickness of the foil ($L_1$) is not crucial for this study. In the simulations we present here we set $L=30\ {\rm \mu m}$, $L_1=4\ {\rm \mu m}$. The dimension of the target in the third ($z$) direction is $30\ {\rm \mu m}$ in 3D simulations.

Due to the limitation of the computational resources, the density of the target is $15n_c$ in 3D simulations, where $n_c=m_e \omega_0^2/4\pi e^2$ is the critical density. In 2D simulations that will be presented later, a higher density up to $100n_c$ is used, which shows little effect on the main findings of this work.
The particle-in-cell (PIC) simulations are carried out with the code EPOCH \cite{Arber2015}. The dimensions of the simulation box are $x \times y\times z=100\lambda_0 \times 80\lambda_0 \times 80\lambda_0$ and are sampled by $2000 \times 800 \times 800$ cells with $8$ macro-particles for electrons.

The laser peak reaches the right end of the target at simulation time $t=167\ {\rm fs}$.
The electron bunches are enveloped in the laser field and pass by the edge of the foil, as shown by the orange dots in Fig.~1(a) at $t =200\ {\rm fs}$, which gives rise to coherent diffraction radiation. The total charge of the beaming electrons is $3.2\ {\rm nC}$. To show the properties of the radiation, we converted to spherical coordinates as $r=\sqrt{(x-x_0)^2+(y-h_2)^2+z^2}$, $\theta={\rm arccos}[(x-x_0)/r]$ and $\phi={\rm arctan}[z/(y-h_2)]$ [illustrated in Fig.1(b)]. The 3D structure of the polar component of the electric fields with frequency below $60\ {\rm THz}$ is shown by rainbow colourscale in Fig.~1(a).
The energy of the polar electric field $E_{\theta}$ together with the azimuthal magnetic field $B_{\phi}$ accounts for about $87\%$ of the total radiation energy, so the THz emission is mainly radially polarized.
The angular distribution of the THz energy is presented in
Fig.~1(c).  The THz emission is predominantly in the forward direction: the total THz energy $W_r$ emitted within $\theta<35^{\circ}$ is $10.7\ {\rm mJ}$, corresponding to an conversion efficiency of $0.83\%$.

\begin{figure}[b]
	\centering
	\includegraphics[width=0.48\textwidth]{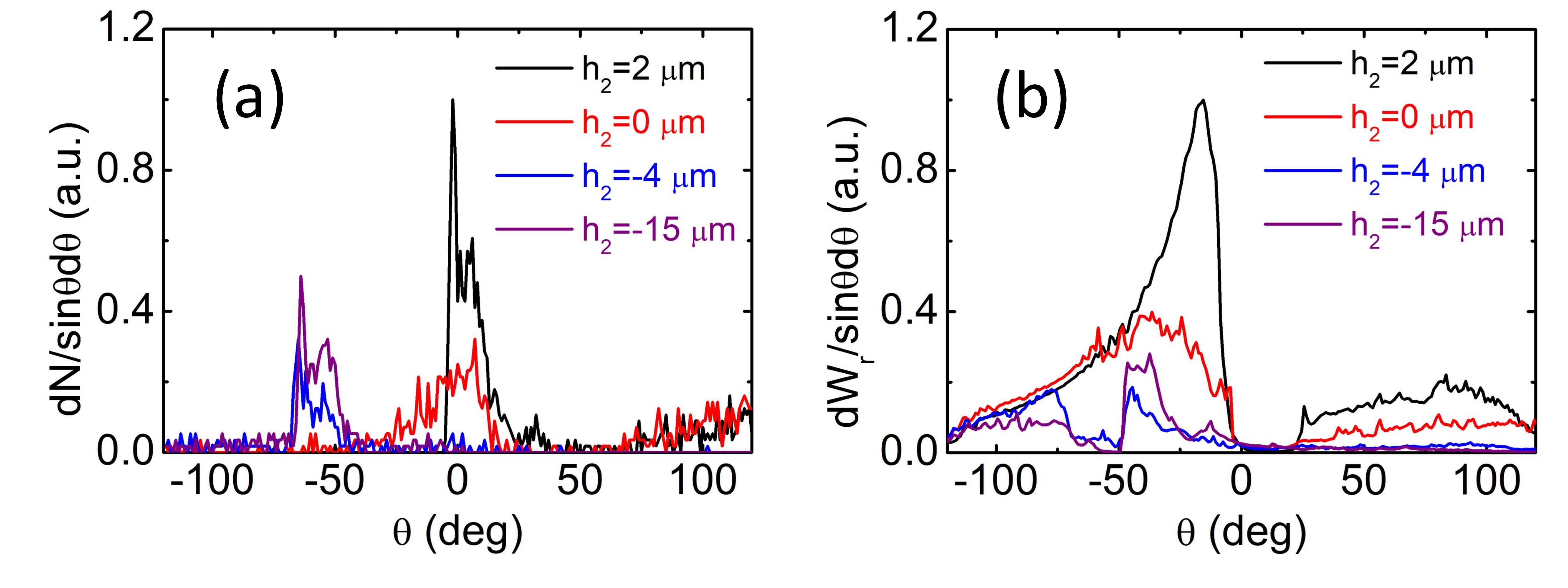}  		
	\caption{
		(a) Angular distribution of fast electrons ($\gamma>5$) in the $x-y$ space at $t=333\ {\rm fs}$. The angle $\theta$ is calculated by
		${\rm arctan}[y/(x-30\ {\rm \mu m})]$.
		(b) Angular distribution of THz energy for different $h_2$.
	} 		
\end{figure}

The temporal structure of the polar component of the electric fields observed at $\theta=5.6^{\circ}$ and $\phi=180^{\circ}$, corresponding to the highest THz energy in Fig.~1(c), is shown in Fig. 1(d). The peak amplitude is $5.5\ {\rm GV/cm}$, corresponding to a normalized amplitude $a_{\rm THz}=2.2$, surpassing the relativistic threshold.
The spectrum of radiation is shown in Fig.~1(e), from which we see that the central frequency is several THz, and over $95\%$ of THz energy in the frequency domain is distributed below $20$ THz.

The property of the electron beams is crucial for determining the energy of radiation, especially a high-charge beam with small divergence is favourable \cite{Schroeder2004, Yi2019}.
We therefore analyze the electron dynamics in the laser solid interaction, using 2D PIC simulations with higher resolution ($dx\times dy=\lambda_0/50\times\lambda_0/50$). The laser and target parameters are the same as in the 3D simulation expect the target density is $100n_c$.
Our scheme is compared with a usual laser-foil interaction setup, where the laser pulse specularly reflects at the front surface of the solid foil (corresponding to $h_2\ll -w_0$ in our setup).

In Figs. 2(a) and (b), we plot the angular distribution of the fast electrons and the corresponding THz radiation based on CTR, respectively. It is found that the displacement of foil right edge and the laser axis ($h_2$) has profound impact on the behaviour of surface electrons.
In particular, when $h_2>0$, significant beaming is observed for the electrons emitted along laser axis (Fig.~2(a)), which leads to enhancement of THz generation via CTR as shown in Fig.~2(b).
For a conventional setup ($h_2=-15\ {\rm \mu m}$), the fast electrons are broadly distributed, with a small bump formed around specular reflection direction due to vacuum laser acceleration \cite{Thevenet2015, Tian2012},
and the strength of THz radiation is much weaker.
Note that the case with $h_2=-4\ {\rm \mu m}$, where most of the incident laser pulse is reflected, shows little difference with the conventional setup, this means the beaming of electron bunches does not depend on the transportation of electrons along the surface, rather it results from the electron energisation and dynamics in the electromagnetic fields near the foil bulk.

To show this, we track the fast electrons ($\gamma>20$, chosen at $t=200\ {\rm fs}$, when the laser peak arrives at the target) throughout the 2D simulations, and plot 20 representative trajectories (randomly chosen) in Figs.~3(a) and (b) for $h_2=2\ {\rm \mu m}$ and $h_2=-15\ {\rm \mu m}$, respectively.
In both cases, surface electrons are initially energised via $J \times B$ heating and vacuum heating to relativistic energies. Their subsequent dynamics is determined by the collective effect of the incident laser and reflected laser pulse \cite{Naumova2004, Li2006,Chen2006}.

\begin{figure}[tb]
	\centering	
	\includegraphics[width=0.48\textwidth]{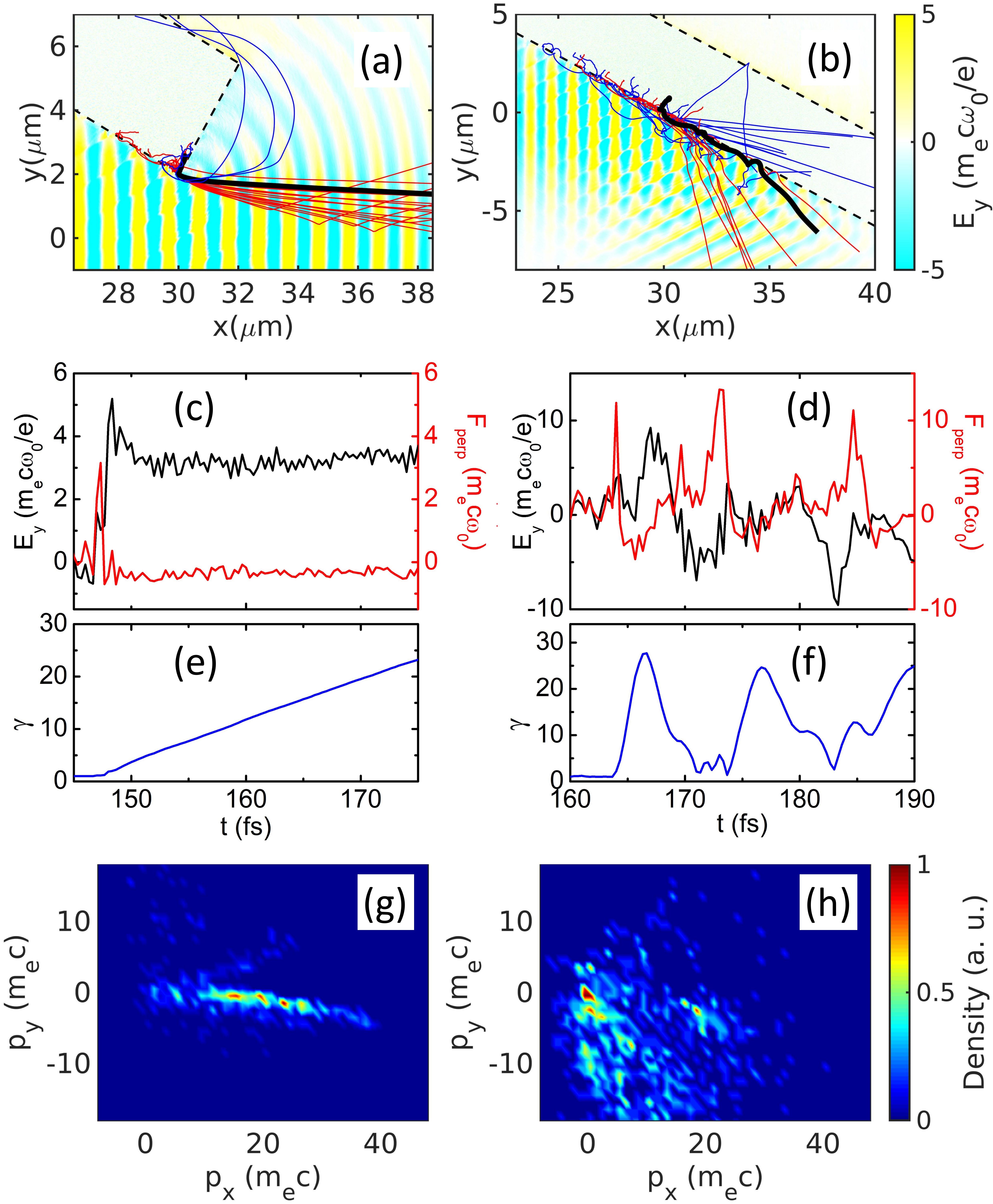}
	\caption{
		(a), (b) Trajectories of $20$ randomly chosen fast electrons ($\gamma>20$ at $t=267\ {\rm fs}$) are displayed, the red and blue lines corresponding to electrons travelling away and towards the target bulk, respectively. the black lines are the representative trajectories for which (c-f) are plotted.
		On the backgrounds are snapshots of $E_y$ at
		$t=167\ {\rm fs}$. Dashed lines mark the initial boundaries of the target.
		(c), (d) Time evolution of normalized electric fields along the $y$ direction $E_y$, as well as the Lorentz forces perpendicular to the electron velocity acting on the representative electrons.
		(e), (f) The $\gamma$ factors versus time for the same electrons in (e) and (d).
		(g), (h) Electron distribution in momentum space $p_x$-$p_y$ at $t=175\ {\rm fs}$.
		Left column is the proposed scheme ($h_2 = 2{\rm \mu m}$) and the right column shows the reference case of conventional laser-foil setup ($h_2 =-15 {\rm \mu m}$).
	} 		
\end{figure}

The beaming effect in our scheme is attributed to direct laser acceleration and the Coulomb force from the target bulk. Since the reflected fields are negligible, the electromagnetic field can be estimated by a plane wave within a few tens of microns (corresponding to the formation length of THz radiation).
The relativistic electrons moving in alignment with its wavenumber $\boldsymbol{k}$ can be locked in a certain phase and experience maximum acceleration, thus most likely to escape from the Coulomb barrier near the foil, as shown by red lines in Fig.~3(a). The electrons travelling with an angle ($\theta_e$) with respect to $\boldsymbol{k}$ experience a dephasing effect, the dephasing length can be estimated by $L_d \approx \lambda_0[2/\cos(\theta_e)-2]^{-1}$. When dephasing happens, the energy of electrons decreases [$d\gamma/dt=-e\boldsymbol{E} \cdot \boldsymbol{\beta}/ (m_ec)< 0$] due to the inverse-sign of $\boldsymbol{E}$, where $\boldsymbol{\beta}$ is the electron velocity normalised by $c$, so they are pulled back by the electrostatic fields to the target bulk (blue curves in Fig.~3(a)). As a result, the electrons traveling above certain critical angle are filtered by the Coulomb force. Assuming the electrons need to travel with the light for a few wavelengths to get sufficient energy, $L_d > \alpha\lambda_0$, where $\alpha$ is on the order of unity, one obtains a beam divergence $\sim 20^{\circ}$ for $\alpha\sim5$, which agrees with the numerical results shown in Fig.~2(a).

In comparison, Fig.~3(b) shows when the laser is totally reflected on the foil, the collective effect of both incident and reflected waves leads to a complex electron motion. Figure~3(c-d) show the time evolution of electric field $E_y$, perpendicular force $F_{perp}$ acting on one representative electron (the trajectory is marked with black in Fig.~2(a-b)) for both cases, whose relativistic gamma factor is shown in Fig.~3(e-f). As one can see, the changing sign of $E_y$ indicates the time when dephasing is happening, which is followed by a sharp reduction in the electron energy. In the mean time the force perpendicular to its velocity $F_{perp}$ increases dramatically, this means the electron is scattered way.

In the conventional laser-foil setup, as the incident and reflected waves travels towards different directions, it is impossible for the electrons to stay in phase with both, the dephasing process happens to all the electrons near the front surface, which results in broad distribution in the momentum space map (Fig.~3(h)). Whereas in the proposed scheme, most electrons in the forward-propagating beam do not experience dephasing, their energy increases monotonically and the $F_{perp}$ remains negligible. These electrons are concentrated along a thin line in the momentum space (Fig.~3(g)), indicating they are traveling with small divergence.

\begin{figure}[t]
	\centering
	\includegraphics[width=0.46\textwidth]{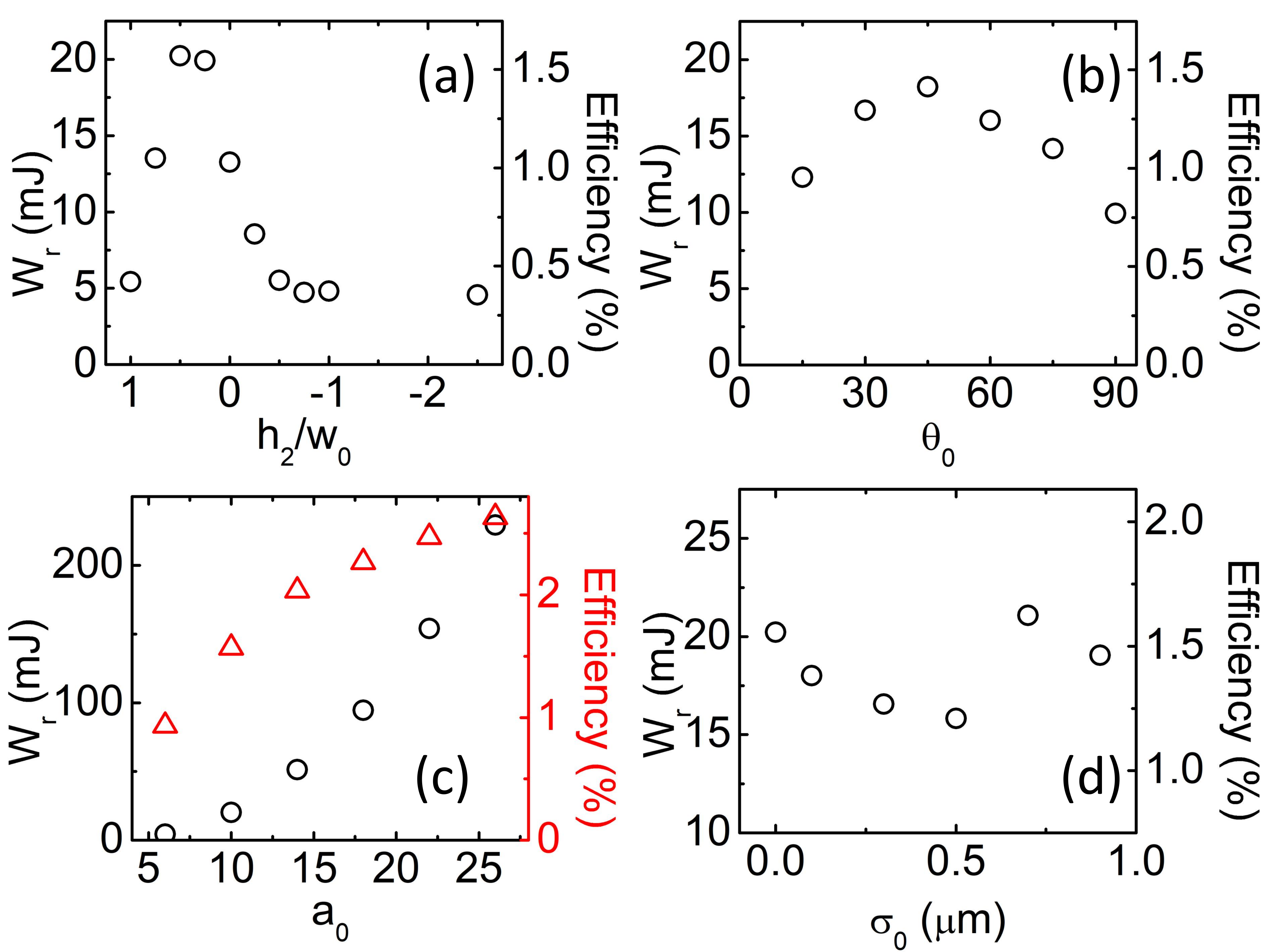}
	\caption{
		THz radiation energy $W_r$ and conversion efficiency are plotted against varying $h_2$ (a), varing laser amplitude $a_0$ (b), varying angle of the target relative to the laser axis $\theta_0$ (c) and varying scale length of the preplasma $\sigma_0$ (d).
		The default parameters are
		$h_2=2\mu$m in (b)(c)(d),
		$a_0=10$ in (a)(c)(d),
		$\theta_0=30^{\circ}$ in (a)(b)(d),
		$\sigma_0=0\ {\rm \mu m}$ (no preplasma) in (a)(b)(c).
	} 		
\end{figure}

In the following, we discuss the dependence of THz energy on laser and target parameters.
In Fig.~4(a), the energy of THz radiation and conversion efficiency are plotted as a function of $h_2$ for a fixed laser amplitude $a_0=10$.
The optimal $h_2$ are $2\ {\rm \mu m}$ and $1\ {\rm \mu m}$, at which the strongest THz radiation with $W_r=20\ {\rm mJ}$ is generated, corresponding to an efficiency of $1.5\%$. The efficiency of forward THz radiation within $-35^{\circ}<\theta<35^{\circ}$ is about $0.9\%$, in reasonable agreement with the 3D simulation result.
When the laser axis is blocked by the target, the mechanism of CTR in the specular reflection direction starts to play the major role, which causes the THz energy decreasing rapidly and finally saturating to $5\ {\rm mJ}$ for $h_2\le-3\ {\rm \mu m}$. The saturated efficiency is smaller than $0.5\%$, on the same level with that in the traditional backward CTR scheme.

We then keep $h_2=2\ {\rm \mu m}$ fixed and varies the initial angle of the target.
For a relativistic femtosecond incident laser, fast electrons are excited most efficiently at $\theta_0=45^{\circ}$  \cite{Gibbon1994}.
This is in accordance with our simulation results presented in Fig.~4(b). The maximum THz energy and efficiency appeared at $\theta_0=45^{\circ}$ are $18\ {\rm mJ}$ and $1.4\%$, respectively. The efficiencies for $30^{\circ}<\theta_0<75^{\circ}$ are maintained over $1\%$.

In Fig.~4(c), the effect of laser intensity is considered.
The efficiency exceeds $1\%$ for $a_0=6$ and increases as the intensity grows. It is also found that at high intensities,
and the efficiency saturates at about $2.5\%$ for $a_0>20$. Thus, the proposed scheme can maintain a high efficiency when scaling towards higher drive laser intensities.

Finally, we consider the pre-expansion due to finite laser contrast by introducing a preplasma on the front surface of the target, $n(d)=100n_c{\rm exp}(-d^2/\sigma_0^2)$.
Here $d$ is the distance perpendicular to the target surface and $\sigma_0$ is the preplasma scale length.
Figure 4(d) shows little variation of the THz energy with the preplasma scale length within a reasonable range, the conversion efficiency is above $1\%$ for all cases. It is evident that our scheme works when a preplasma exists.
Note that the pre-expansion leads to an effective surface of the target at where $n(d_1)=n_c$. In this case, $h_2$ is measured as the distance between the right end of the effective surface and the laser axis.

In conclusion, we have proposed a novel scheme on generating high-intensity, well-collimated THz radiation via interactions between a femtosecond, relativistic incident laser and a solid foil target.
When the foil does not cover the laser axis and the reflected pulse is weak, a group of surface electrons are dragged out and accelerated along the direction of laser propagation by the laser field.
A substantial portion of electron energy is transferred to an intense THz radiation pulse due to coherent diffraction radiation, when the beaming electrons pass by the target edge.
According to 2D and 3D PIC simulation results, the generated THz energy can reach $20\ {\rm mJ}$ and the pulse can be fully relativistic. The conversion efficiency can be over $1\%$. Compared with traditional THz sources based on laser-solid interactions, the conversion efficiency is several times higher.
\section*{Acknowledgements}
The authors would like to thank Prof.~T F\"ul\"op for fruitful discussions.
This work is supported by the Olle Engqvist Foundation and has received funding from the European
Research Council (ERC) under the European Union's Horizon 2020 research and
innovation programme under grant agreement no.~647121. Simulations were performed on resources at Chalmers Centre for Computational Science and Engineering (C3SE) provided by the Swedish National Infrastructure for Computing (SNIC).

\end{document}